![symmetry logo] ![MDPI logo]

*Article*

# AntiPhishStack: LSTM-based Stacked Generalization Model for Optimized Phishing URL Detection

Saba Aslam [1,2], Hafsa Aslam [3], Arslan Manzoor [4], Chen Hui [5,*], Abdur Rasool [1,*]

1. Shenzhen Institute of Advanced Technology, Chinese Academy of Sciences, Shenzhen 518055, China
2. Shenzhen College of Advanced Technology, University of Chinese Academy of Sciences, Shenzhen 518055, China
3. Dibris Polytechnic Interschool Section, University of Genoa, Genoa, Italy
4. Department of Mathematics and Informatics, University of Catania, Catania, Italy
5. Shenzhen Polytechnic University, Shenzhen, 518055, Guangdong, China
* Correspondence: hui.chen1@siat.ac.cn (C.H.), rasool@siat.ac.cn (A.R.)

**Abstract:** The escalating reliance on revolutionary online web services has introduced heightened security risks, with persistent challenges posed by phishing despite extensive security measures. Traditional phishing systems, reliant on machine learning and manual features, struggle with evolving tactics. Recent advances in deep learning offer promising avenues for tackling novel phishing challenges and malicious URL. This paper introduces a two-phase stack generalized model named AntiPhishStack, designed to detect phishing sites. The model leverages the learning of URL and character-level TF-IDF features symmetrically, enhancing its ability to combat emerging phishing threats. In Phase I, features are trained on a base machine learning classifier, employing K-fold cross-validation for robust mean prediction. Phase II employs a two-layered stacked-based LSTM network with five adaptive optimizers for dynamic compilation, ensuring premier prediction on these features. Additionally, the symmetrical predictions from both phases are optimized and integrated to train a meta-XGBoost classifier, contributing to a final robust prediction. The significance of this work lies in advancing phishing detection with AntiPhishStack, operating without prior phishing-specific feature knowledge. Experimental validation on two benchmark datasets, comprising benign and phishing or malicious URL, demonstrates the model's exceptional performance, achieving a notable 96.04% accuracy compared to existing studies. This research adds value to the ongoing discourse on symmetry and asymmetry in information security and provides a forward-thinking solution for enhancing network security in the face of evolving cyber threats.

## 1. Introduction

Phishing, a deceptive method through social and technical engineering, poses a severe threat to online security, aiming to obtain illicit user identities, personal account details, and bank credentials [1]. It's a primary concern within criminal activity, with phishers pursuing objectives such as selling stolen identities, extracting cash, exploiting vulnerabilities, or deriving financial gains [2, 3]. The nuanced landscape of phishing techniques showcasing symmetry and asymmetry includes algorithms, domain spoofing, HTTPS phishing, SMS phishing, link handling, email phishing, and pop-ups. Attributes such as prefixes, suffixes, subdomain, IP address, URL length, '@' symbol, spear phishing, dual-slash attributes, port, HTTPS token, request URL, URL-anchor, tag-links, and domain age contribute to the multifaceted nature of phishing attacks [4]. Phishing perpetrators adeptly mimic legitimate websites, particularly those related to online banking and e-commerce. This creates a symmetrical illusion that induces users to unwittingly divulge sensitive information, leading to various fraudulent actions [5, 6].

A phishing attacker's role involves three specific duties: influencing target selection, sociological aspects, and technological infiltration [7]. As of March 2006, the Anti-Phishing Working Organization reported 18,480 significant phishing assaults and 9666 distinct phishing domains, resulting in substantial financial repercussions for businesses and affecting billions of site visitors [8]. Microsoft estimates the potential cost of computerized offenses on the global network to be a staggering 500 billion USD, underscoring the symmetrical impact of cyber threats on the financial ecosystem [9]. A single data breach could incur an average cost of approximately 3.8 million USD for organizations in 2018, highlighting the symmetrical consequences of security lapses. Data from the Anti-Phishing Working Group (APWG) reveals a notable increase in attack networks, with 180,768 identified during the first quarters of 2019, up from 138,328 in the fourth quarter of 2018 and 151,014 in the third quarter of 2018 [10]. The visual symmetry between benign and deceptive websites challenges human perception, making it difficult to distinguish between them. When visitors access these mimicked sites, critical information is stolen through scripting, underscoring the symmetrical vulnerability in human-computer interaction. The exponential growth in e-commerce consumers contributes to the escalating frequency of phishing attacks, carried out through various means such as malware, online platforms, and email, creating a symmetrical escalation in cyber threats [11].

Researchers propose varied solutions to enhance symmetry in phishing detection. Some use a blacklist for identifying phishing sites [12]. However, this method fails to detect non-blacklisted phishing websites, introducing asymmetry, such as zero-day attacks. Heuristic-based detection analyzes website content and third-party service features, but potential service restrictions create asymmetry. Simultaneously, exploring online content and third-party features introduces temporal asymmetry due to its time-consuming nature [13]. Similarly, a hierarchical clustering method groups DOM vectors based on distance, limiting detection efficiency and suggesting a need for symmetrical analysis of URL features to enhance throughput [14].

URL play a pivotal role in phishing attacks, transmitted to users through various channels like emails and social media, presenting a facade of symmetry by appearing as genuine URL [15]. Machine learning-based techniques emerge as symmetrical solutions among the available approaches for evaluating URL. By familiarizing malicious URLs with categorization algorithms, these techniques effectively differentiate between phishing and benign URL, introducing a symmetrical balance in the categorization process [16]. URL-based studies leverage a phishing tank database, a comprehensive collection tracking reported phishing URL by various online security companies. While this database offers organized data categorization patterns, asymmetries arise when using categorization algorithms or machine learning for URL data, necessitating additional symmetrical URL management techniques [17]. Standard techniques like blacklisting, regular expression, and signature matching, although employed to identify phishing attempts, exhibit asymmetry by falling short in detecting unfamiliar URL [4]. Continuous updating of database signatures to detect unexpected patterns in malicious URL underscores the need for applying symmetrical machine learning-based research, particularly with deep learning models, for robust and symmetrical identification of malicious URL [18].

Machine learning and deep neural networks have been pivotal in various research endeavors, showcasing substantial performance improvements [19-22]. In the context of phishing detection, authors [19] proposed a multidimensional feature engineering approach, harnessing a deep learning model (CNN-LSTM) and machine learning algorithms. This method integrated predictions using the XGBoost (eX-treme Gradient Boosting) algorithm, offering a solution to extract features from diverse dimensions for swiftly effective attack detection. However, the reported results indicated a decline in the false positive rate to 59%, signaling a reduction in the level of attack prediction. Another study [20] introduced an end-to-end deep learning architecture grounded in natural language processing techniques to combat malicious URL phishing. The model aimed to classify benign and malicious URL using character-level and word-level embedding in CNN networks. However, the model exhibited a lack of generalization on test data, indicating a need for improved accuracy and malicious URL detection ability. Wang et al. [21] presented the PDRCNN approach, designed to enhance phishing detection efficiency by eliminating reliance on feature crawling from third-party services. Based on the LSTM network, this approach selects optimal features from the URL, employs CNN to distinguish characters influencing phishing, and ensembles predictions with machine learning classifiers. While reporting efficient performance, the mechanism's dependency on existing knowledge of phishing detection raises concerns about its susceptibility to errors in identifying the latest vulnerabilities.

In contrast to traditional machine learning methods that implicitly extract handcrafted features, deep learning approaches prove advantageous when faced with the challenge of professional phishers exploiting the multilayer features of URLs. To address this, stacking, an ensemble learning methodology integrating various machine learning algorithms and deep learning models, employs a metamodel to amalgamate predictions, enhancing overall performance. Initially employed for malware identification on mobile devices, the stacking approach demonstrated improved accuracy and F measure [23]. We have extended this stacking mechanism by designing two distinct phases, leveraging the symmetrical integration of other methods to enhance detection impact.

This paper leverages a deep-learning neural network, long-short-term memory (LSTM), introducing a novel stack generalization model named AntiPhishStack. The proposed model employs five optimizers in two phases to detect phishing URLs effectively. In the first phase, machine learning classifiers, coupled with K-fold cross-validation to mitigate overfitting, generate a mean prediction. The second phase utilizes a two-layered LSTM-based stack generalized model optimized for premier prediction in phishing site detection. Merging the mean prediction from Phase I with the premier prediction from Phase II, meta-classifiers, specifically XGBoost, deliver the final prediction. This stacking model significantly enhances phishing detection accuracy by learning URL and character-level TF-IDF features, showing symmetrical capabilities. The AntiPhishStack model intelligently identifies new phishing URLs previously unidentified as fraudulent. Experimental evaluations on two benchmark datasets ([24] and [25]) for benign and phishing sites demonstrate robust performance, assessed through various matrices, including AUC-ROC curve, Precision, Recall, F1, mean absolute error (MAE), mean square error (MSE), and accuracy. Comparative analysis with baseline models and traditional machine learning algorithms, such as support vector machine, decision tree, naïve Bayes, logistic regression, K-nearest neighbor, and sequential minimal optimization, highlights the AntiPhishStack model's superior phishing detection efficiency. Notably, this model offers the following significant advantages in achieving symmetrical advancements in cybersecurity:

i. *Prior feature knowledge independence*: The approach taken in this work embraces the concept of symmetry by treating URL strings as character sequences, serving as natural features that require no prior feature knowledge for our proposed model to learn effectively.

ii. *Strong generalization ability*: The URL character-based features are utilized for more robust generalization and check-side accuracy, and the multi-level or low-level features are combined in the hidden layers of the neural network to attain effective generalization.

iii. *Independence of cybersecurity experts and third-party services*: Our proposed stack generalization model autonomously extracts necessary URL features, eliminating the reliance on cybersecurity experts. Additionally, the AntiPhishStack model, reliant on URL and character-level TF-IDF features, demonstrates independence from third-party features such as page rank or domain age.

The significant contributions of this paper are:

- Presentation of a two-phase stacked-based generalization model (AntiPhishStack) that breaks free from the necessity of prior feature knowledge for phishing site detection. The model achieves this by learning URL and character-level TF-IDF features.
- In Phase I, features are trained on the base machine learning classifier to generate the mean prediction. Meanwhile, Phase II employs two-layered stacked-based LSTM networks and five adaptive optimizers for premier prediction detection.
- The final prediction is established by developing a meta-classifier (XGBoost) classifying URL into benign and phishing categories. Experimental results showcase the AntiPhishStack model's noteworthy performance on baseline models, utilizing symmetrically structured Alexa and PhishTank datasets.

The structure of the rest of the article is as follows: Section 2 deliberates the background research work of phishing detection, Section 3 introduces the AntiPhishStack proposed model, Section 4 delivers the experiments, and Section 5 presents the results and its evaluations, and Section 6 elaborates the conclusion and future work.

## 2. Literature Review

### 2.1. Phishing detection

There are three different methods for phishing detection which are often utilized [26]. First and foremost, web-based phishing refers to imitating a legitimate web interface. Phishers trick users into providing credential information, believing it to be genuine. Second, attackers transmit phishing material via email using web-based methods. The third is a malware-based phishing assault in which attackers insert harmful code into the user's system [16]. Adebowale [27] suggested a common approach in which particular users steal private information from websites and are called phishing users. This behavior is usually carried out through phony websites or malicious URLs, which are called fraudulent enterprises. Cybercriminals develop a well-planned phishing assault by engaging in fraudulent activities. After gaining access to the victim's computers, hackers may install malware or insufficiently secure users' systems. Acquisti [28] proposed that several techniques are advised to prepare and teach end-users to detect phishing URLs to decrease the potential of phishing attacks. El-Alfy [29] suggested using the architecture of the node to train unsupervised and supervised algorithms. Phishing sites rely on feasibility neural networks and K medoids clustering. The k-medoid method uses feature selection and modules to minimize storage capacities. The required technique achieves 96.79% accuracy on thirty characteristics. PHISH-SAFE, an anti-phishing solution, has suggested employing an SVM classifier to recognize phishing websites with greater than 90% accuracy [30].

One research offered another anti-phishing approach based on a weighted URL tokens system that extracts identification keywords from a query webpage. A search engine was used to locate the target domain name by identifying keywords as search phrases and validating the query web page. Tan et al. presented anti-phishing approach that collects keywords from a website and then uses a weighted URL tokens-based system [31].

### 2.2. Machine learning-based detection

In the last couple of decades, machine learning has been vigorously applied for phishing detection through different models and methods for various purposes [32-34]. For instance, Wang [35] proposed ensemble classifiers for email filtering that eliminated five algorithms: Support Vector Machines, K-Nearest Neighbor, Gaussian Naive Bayes, Bernoulli Naive Bayes, and Random Forest Classifier. Finally, random forest enhanced its accuracy from 94.09 percent to 98.02 percent.

Table 1. The literature review summary compares phishing detection studies utilizing machine and deep learning techniques.

| C | Ref. | Method | Datasets | Findings | Limitations and future gaps |
|---|---|---|---|---|---|
| Phishing detection | [36] | Character embedding CNN and RF to classify phishing websites based on multi-level features. | PhishTank (47,210) and Alex (83,857) | 95.49% on dataset D2 | Reliance on URL features only and limited to character-level features. |
| | [29] | PNN and K-medoids clustering are combined and trained on pre-classified websites and relevant features. | 11,055 phishing and benign websites | 87% using address bar-related features | The potential impact of the Gaussian smoothing parameter on performance and the limited effectiveness of HTML- and JavaScript-based features. |
| | [37] | Two-level filtering mechanism using lightweight visual similarity and heuristic filtering to detect phishing. | PhishTank and Google, 100 search results | Matthew's correlation coefficient is 97% | Inability to detect variations of blacklisted sites and the generation of false negatives when encountering out-of-list legitimate sites. |
| ML-based detection | [38] | Rule-based system that extracts hidden knowledge to detect phishing attacks. | Dataset3 (103 phishing and 73 legitimate) | Average accuracy 90.51% with SVM and 1.35% error rate | Reliance on webpage content for feature sets may not account for attackers redesigning phishing web pages. |
| | [24] | Whitelists and blacklists for classifying legitimate and phishing web pages. | Ebbu2017 contains 73,575 URLs | Accuracy rate of 10.86% with DT | Limited dataset size (1400 items) and the high acceptance for noisy data. |
| | [25] | Detects phishing sites using client-side, URL-based features, independent of third-party services, and fast computation. | Common-crawl, Alexa database, PhishTank, total samples: 85,409 | CatchPhish accuracy is 94.26% with Random Forest | Model misclassified some phishing sites hosted on free or compromised hosting servers. |
| DL-based detection | [39] | Mapas detects malware using API call graph patterns with a CNN model. | 9,000 malicious apps, 9,000 benign from Google Play | Accuracy 91.27% with CNN | Excludes obfuscated apps that cannot extract API call graphs with flow-droid. |
| | [21] | Uses RNN to extract global features and CNN for local features from URLs for phishing detection. | Alexa and PhishTank, 500,000 sample | Accuracy 93.48% (RNN) 95.03% (CNN) | Training time was too long, and unable to classify URLs if it's not semantics. |
| | [19] | Utilizes a CNN-LSTM with feature extraction for phishing detection. | PhishTank and dmoztools.net (989,021 URLs) | Accuracy is 94.41% | It is not supported for webpage code and webpage text detection. |

Rahman et al. [40] utilized six machine learning classifiers (KNN, DT, SVM, RF, ERT, and GBT). They applied three publicly accessible datasets with multidimensional attributes that could also be used to detect phishing attacks in several anti-phishing systems due to a lack of proper selection of machine learning classifiers. To quantify the classifier performance, use the confusion matrix, precision, recall, F1-score, accuracy, and misclassification rate. It finds greater performance than Random Forest and Exceptionally Randomized Tree, which achieved 97% and 98% accuracy rates for detecting phishing URLs, respectively. Gradient Boosting Tree performs best for the multiclass feature set, with 92% accuracy.

Mogimi et al. [38] proposed a phishing detector that exhibits a lack of symmetry in its approach. The support vector machine (SVM) method is initially employed to train a phishing detection model, followed by the decision tree (DT) approach to uncover concealed phishing. Although the suggested method achieves high true positive rates (0.99) and low false negative rates (0.001) in a large dataset, it operates under the assumption that phishing web pages exclusively utilize innocuous page content. This assumption lacks symmetry with real-world scenarios, where phishing pages may employ deceptive elements. Rao et al. [25] recently presented CatchPhish, a lightweight program that predicts URL validity without examining the website's content. The suggested framework uses the random forest classifier to retrieve the suspicious URL's hand-crafted and Term Frequency-Inverse Document Frequency (TF-IDF) characteristics.

*2.3. Deep learning-based detection*

Machine learning and deep learning are both types of AI. Machine learning, in essence, is AI that uses algorithms to read data, learn from that data, and make decisions based on what it has learned, allowing it to adapt automatically with little to no human intervention. However, deep learning uses artificial neural networks, a specialized branch of machine learning, to simulate how the human brain learns. It layers algorithms to build an "artificial neural network" to learn and decide for itself.

In phishing detection, deep learning provides an automatic, accurate, and fast means (artificial neural network) to identify the URLs as Benign or legitimate. It uses different layers and weights assigned by the optimization functions to learn the huge complex dataset, which traditional machine learning algorithms find difficult to process. The development of deep learning algorithms, e.g., recurrent neural networks (RNN), recurrent convolutional neural networks, and deep neural networks (DNN), have lately been used for phishing detection. Though deep learning approaches are not often used in phishing detection due to the lengthy training period, they frequently give higher accuracy and automatically retrieve characteristics from raw data with no background experience [11, 41]. Neural networks typically include one to two hidden layers. The number of layers varies in various deep-learning applications. However, it needs almost 150 layers [20]. There are several guidelines for determining the number of layers, including two or fewer layers for basic data sets and additional layers for computer vision, time series, or complicated datasets [4, 39, 42].

Wang et al. [21] presented a rapid phishing website detection approach dubbed accurate phishing detection with recurrent convolutional neural networks (PDRCNN) based solely on the website's URL. It converts URL information into a two-dimensional tensor and feeds the tensor to a deep-learning neural network to categorize the original URL. They first employ a bidirectional long short-term memory (LSTM) network to extract global and local URL characteristics, followed by a convolutional neural network (CNN). YANG et al. [19] created a two-step multidimensional feature phishing detection technique. The character sequence characteristics of the provided URL are retrieved and utilized for categorization by LSTM-CNN deep learning networks in the first phase. In the second phase, URL statistics data, webpage content characteristics, and deep learning classification results are merged to form multidimensional features. Yuan et al. [43] Yuan et al. [36] developed a method introducing symmetry by combining character embedding (word2vec) with URL structure to create vector representations of URLs. The URL is systematically divided into five components: the URL protocol, sub-domain name, domain name, domain suffix, and URL path. This approach enhances existing classification methods, training vector representations to identify phishing URLs symmetrically. Huang et al. [1] presented a deep learning-based approach for detecting phishing URLs called PhishingNet. They utilize a CNN network to extract character-level URL characteristics, and an attention-based hierarchical recurrent neural network (RNN) to retrieve word-level URL features. The features are then fused and trained using three convolutional layers and two fully connected layers.

The summary of these studies is categorized (C) in Table 1.

*2.4. Stack generalization-based detection*

Deep learning learns representation at several levels of abstraction by using layers of layered nonlinear projections. It has demonstrated superior performance in various applications, including natural language processing, computer vision, speech recognition, and so on [44-47].

The DNN-recommended [20] are trained with inferred deep stacking. The analyzed covers of previous outlines are updated as they had been at the end of each DNN training epoch, and the upgraded evaluated veils then provide further inputs to train the DNN in the subsequent epoch. During the testing phase, the DNN generates expectations sequentially and repeatedly. In addition, it suggests using the L1 loss for training.

## 3. AntiPhishStack Proposed Model

The primary purpose of this model is to determine the best output through evaluation by applying the stacking technique and deep neural network to the processed data set and to propose an optimized model based on that output. The notations and meanings used in this paper are described in Table 2. The AntiPhishStack model of stack generalization has been illustrated in Fig. 1.

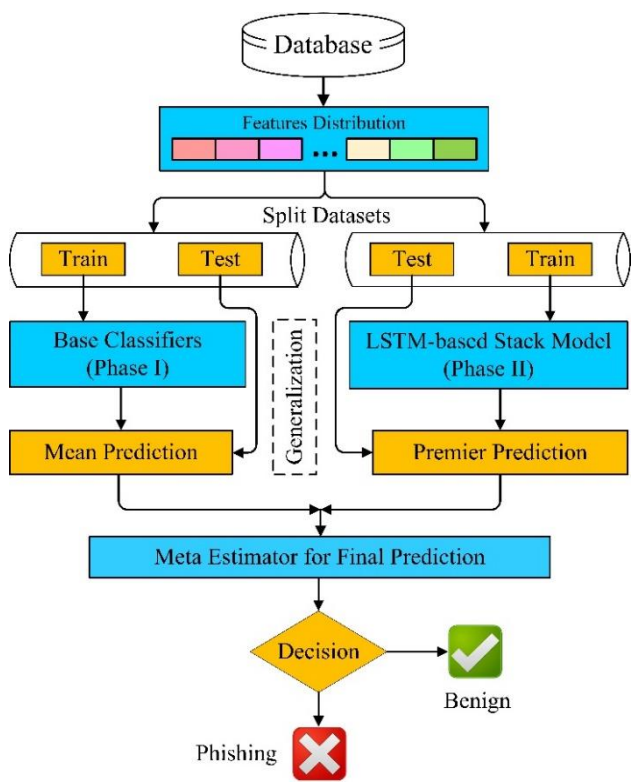

**Figure 1.** AntiPhishStack: proposed LSTM-based stock generalization model's flow.

Our model's flow has five level approaches. The key steps are as follows:
i. Collection of datasets and feature distribution into URL features and character-level features.
ii. Dataset division into training and testing by 70:30 ratio, respectively.
iii. Construct the stack generalization model's first phase (Phase I) based on the machine learning base model and calculate the mean prediction with the test dataset.
iv. Construct the second phase (Phase II) of the stack generalization model with the LSTM model based on adaptive optimizers and compute the performance evaluation with the test set.
v. Merge predictions and evaluations from both Phase I and Phase II for the ultimate prediction, enhancing symmetrically the recognition and determination of phishing web pages.

**Table 2.** Notations and their meanings.

| Notations | Meaning |
|---|---|
| $W_i$ | Weight factor of URLs |
| $h_{t-1}$ | Hidden state of the $t-1$ instant |
| bb | Bias of each gate |
| $i_t, f_t, o_t$ and $C_t$ | Input gate, forget gate, output gate, and unit status, respectively. |
| $W_f, W_i$ and $W_o$ | Weight matrix of forget gate, input gate, and output gate, respectively. |
| $x_t$ | Current input |
| $f_l$ | Training loss function |
| $\gamma$ | Complexity of each leaf |
| $T$ | Number of leaves nodes |

*3.1. Datasets*

The URLs are collected from a variety of sources (Alexa and PhishTank) [25] [24]. URLs that were duplicated or did not survive were deleted before they were used to create a dataset. The typical URL elements, such as "http://", "https://", and "www." are deleted. Inconsistent URL forms can easily impair the model's quality during training if the prefixes are not trimmed. The database management system (pgAdmin) was utilized in conjunction with Python to import the preprocessed data, and then the dataset was divided into two parts: 70% for training and 30% for testing. The distribution of legitimate and phishing URLs are as follows:
i. *Dataset 1 (DS1)*: Benign Yandex sites (https://yandex.com.tr/dev/xml/) and PhishTank phishing sites [24].
ii. *Dataset 2 (DS2)*: Benign sites from common-crawl, the Alexa database, and phishing sites from PhishTank [25].

The datasets were selected for their diverse and current mix of benign and phishing URLs, ensuring robust model training. DS1 and DS2 offer a balanced representation of typical internet environments and specialized sources, respectively. This variety enhances the model's applicability and accuracy in real-world phishing detection. Meanwhile, the feature dataset is divided into 70% training and 30% testing datasets to ensure a balanced setup: 70% for training our AntiPhishStack model and 30% for robust testing on unseen data, aligning with standard machine learning practices.

*3.2. Feature distribution*

Features and the capacity to use these features must be examined before examining the features selection section [48]. There are four major features and a total of 30 sub-features. Based on the details, each characteristic provides information on whether the website is phishing, legitimate, or suspect. This section contains the plans for highlighting the characteristics.

*3.2.1. URL Features*

Uniform Resource Locator (URL) provides the location of online resources such as pictures, files, hypertext, video, etc. In general, attackers attempt to build phishing URLs that look to users as reputable websites. Attackers use URL jamming tactics to mislead users into disclosing personal information that can be exploited against them. This research aims to detect phishing websites quickly, utilizing lightweight characteristics, i.e., weight factor URL token system, inspired by [49]. For example, the segmentation of URL (Fig. 2) provides the different tokens and their final weight $W_i$ for $i$-th distinct words can be calculated as:

$$W_i = \frac{h_i}{n} \sum_{x=1}^{S} \frac{N_x}{x^2} \quad (1)$$

where, $h_i$ indicates the length of $i$-th distinct word, $S$ denotes the total steps available for tokens, $n$ shows the number of URLs from webpages, $N_x$ total number of $i$-th word occurrences in step $S$ with respect $x$ level.

Calculating this weight delivers the weight value of each URL assigned to neural network gates for phishing prediction. This is accomplished by extracting only characteristics from the URL rather than accessing the website's content. Fig. 2 shows an example of URL characteristics for the weight.

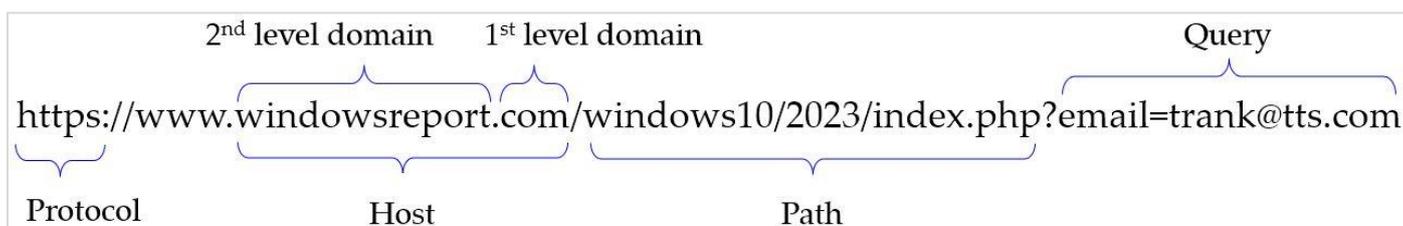

**Figure 2.** Tokenization of URL characteristics and components for the weight calculation.

The first component of the URL is a protocol (https, http, ftp, etc.), which is a set of rules that regulates how data is transported from data transmission. The second component is the location of the host IP address or resource. The hostname is separated into two parts: major domains and top-level domains (TLD). The URL's hostname is comprised of the principal domain and the TLD. The hostname is followed by a port number, which is optional. The third component uses the path to identify the specific resource inside the domain accessed by a user. An optional field, such as inquiry, follows the path. The protocol, hostname, and URL path are appended to the base URL. The combination of the second domain and top-level domain names, known as the host domain, makes the URL unique. As a result, cybersecurity firms are working hard to identify the fraudster websites used for phishing offenses by name. If a hostname is designated as phishing, an IP address can be banned to prevent it from accessing the web pages included within it.

It has the following sub-features, according to the dataset:

- **IP Address**: If an IP address is used instead of a domain name in the URL of a phishing website, the client may virtually be certain that someone is attempting to steal his credentials. From this dataset, 570 URLs with an IP address were discovered, accounting for 22.8 percent of the dataset, and a rule IP address is in URL that is termed Phishing; otherwise, it is Legitimate was suggested.
- **Operate the @ Symbol**: Web browsers usually ignore the section preceded by the @ sign. Because it is maintained separately from real-world addresses, finding 90 URLs with the '@' sign will provide just 3.6 percent of the total, according to the dataset.
- **Operate the "//" symbol**: As valid URLs, the "//" sign is used after HTTP or HTTPS. If the URL changes after the initial protocol declaration, it is called a phishing URL. The "//" sign is used to redirect to other websites.
- **Domain name prefixes and suffixes separated by the "-" sign**: A URL with the "-" sign in its domain name is a phishing URL. In general, verified URLs do not include the "-" sign.
- **Use the "." sign in the domain**: Use the "." sign in the domain. Adding a sub-domain with the domain name must include the dot. Consider it suspect if you drop out more than one subdomain, and anything greater than that will indicate phishing.
- **HTTPS (secure socket layer)**: The majority of legal sites use the HTTPS protocol. Therefore, the age of the certificate is quite important when utilizing HTTPS. This necessitates the use of a trustworthy certificate.
- **Favicon**: A favicon might redirect clients to dubious sites when layered from outside space. It is mainly used on websites and is a graphic picture.

*3.2.2. Character Level Features*

Term Frequency-Inverse Document Frequency is abbreviated as TF-IDF. The TF-IDF score indicates a term's relative significance in the document and throughout the whole corpus. The TF-IDF score is made up of two terms: the first computes the normalized Term Frequency (TF), and the second computes the Inverse Document Frequency (IDF), which is calculated as the logarithm of the number of documents in the corpus divided by the number of documents in which the specific term appears [50, 51].

$$TF(t, d) = \frac{Number\ of\ times\ term\ t\ appears\ in\ a\ document\ d}{Total\ number\ of\ terms\ in\ the\ document} \quad (2)$$

$$IDF(t, D) = log_e(\frac{Total\ number\ of\ documents\ D}{Number\ of\ documents\ with\ term\ t\ in\ it}) \quad (3)$$

$$TF - IDF(t, d, D) = TF(t, d) * IDF(t, D) \quad (4)$$

TF-IDF Vectors may be produced at many levels of input tokens (words, characters, n-grams):

- **Word level TF-IDF:** A matrix indicating the TF-IDF scores of each term in distinct texts.
- **Character level TF-IDF:** A matrix indicating the TF-IDF scores of character-level n-grams in the corpus.
- **N-gram level TF-IDF:** N-grams are the collection of N terms. This matrix indicates the TF-IDF scores of N-grams.

It should be mentioned that TF-IDF has been used in numerous studies to identify website phishing by examining URLs [25] to get indirectly related connections, target websites, and the validity of suspicious websites [50]. TF-IDF retrieves prominent keywords from the textual content. However, it has certain limitations. One of the limitations is that the approach fails when extracting misspelled terms. Because the URL might contain nonsensical words, it used a character-level TF-IDF method with a maximum feature count of 5000.

Furthermore, we have measured the URL strings as character sequences by employing the idea from the literature [52]. This idea provides the advantage that the proposed model can train the URL character sequences as natural features that do not need prior feature knowledge to be learned by our proposed model. Our proposed AntiPhishStack model uses the stack generalization model to extract the local URL features from the URL character sequences. Finally, the URL will be classified by designing a meta-classifier for final prediction.

*3.3. Stack generalization model*

The stack generalization model is divided into two phases, as illustrated in the flow model (Fig. 1).

*3.3.1. Phase I*

Based on the abovementioned characteristics, existing machine learning models are utilized directly to distinguish phishing and legitimate web pages. This paper proposes a stacking model (illustrated in Fig. 3) for this purpose by merging various machine learning models, including support vector machine (SVM), naïve Bayes (NB), decision tree (DT), logistic regression (LR), K-nearest neighbors (KNN), sequential minimal optimization (SMO), and XGBoost.

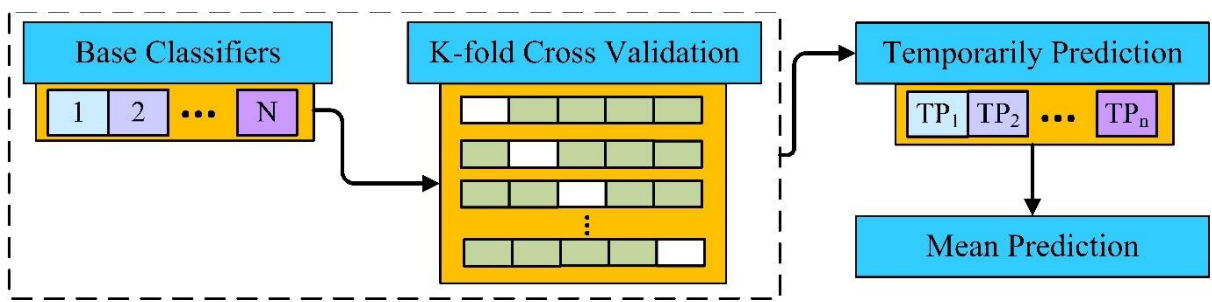

**Figure 3.** Phase I of the proposed stack generalization model.

The training set is split into $Z$ copies, with $Z-1$ copies utilized for training and one copy used for testing. The training process will not be terminated until each basic model has predicted the samples. This suggested system employed k-fold cross-validation to avoid overfitting for this training set and each fold of the train part that might be predicted using out-of-fold.

This suggested model uses a value of three to ten for k-fold cross-validation; after all, it delivers output using a test set. Following the temporary prediction (TP) acquisition, the mean prediction is obtained, which is strengthened by the test dataset validation. This time, it is comprehensive, with a fold approach required for estimating all figures on all folds utilized.

*3.3.2. Phase II*

The train segment is put to a two-layer neural network architecture of LSTM once the features from the training dataset have been loaded. Because there are dependencies on immediately preceding entries in sequential phishing webpage data, LSTM is better suited to simulate phishing detection in this investigation. Meanwhile, it is explicitly designed to avoid the long-term dependency problem by storing the feature information in its memory cell. It can remove or add information to these call states and is regulated by structures called gates. These gates and corresponding operations/functions are presented in [53], while Phase II of the integrated stack generalized model is illustrated in Fig. 4.

In the first gate (Forget gate), the information from the current input $x_t$ and the previous hidden state $h_t$ is passed through the sigmoid activation function. If the output value of the feature is closer to 0, it means forget, and closer to 1 means to retain. The second gate, the input gate, decides what relevant feature (phishing or benign) can be added from the current step. The third gate, the control gate, decides which values will be updated (either 0 or 1), for which a $tanh$ layer creates a vector of $\hat{C}_t$. The last gate, the output gate, determines the value of the next hidden state [54].

At time $t$, the LSTM cell's components are modified as follows:

i. Equation (5) represents the forgotten gate $f_t$ with the sigmoid function $\sigma$. The weights $W_f$ and bias $b_f$ are applied to the concatenation of the previous layer's output $h_{t-1}$ and the current layer's input $x_t$, represented as $[h_{t-1}, x_t]$. This concatenation forms a row vector, and Eq. (5) describes the computation for $f_t$, considering the forgotten information from the cell state at time $t-1$.

$$f_t = \sigma(W_f \cdot [h_{t-1}, x_t] + b_f) \quad (5)$$

ii. Save information in the cell state, which consists mostly of three parts:

a. The Sigmoid layer's results are $i_t$ entering the gate as information to be updated;

b. The $tanh$ layer's freshly generated vector $C_t$ is being added to the cell state. The previous cell state $C_{t-1}$ is multiplied by $f_t$ to forget the information and the new party information $i_t * \hat{C}_t$ is totaled to create a cell state update.

$$i_t = \sigma(W_i \cdot [h_{t-1}, x_t] + b_i) \quad (6)$$

$$\hat{C}_t = tanh(W_c \cdot [h_{t-1}, x_t] + b_c) \quad (7)$$

$$C_t = f_t * C_{t-1} + i_t * \hat{C}_t \quad (8)$$

c. The output gate decides the output data. To process the cell state, first, the Sigmoid layer is used to identify which part of the information should be produced, and then use $tanh$ to process the cell state. The output value is the product of the two elements of the information.

$$o_t = \sigma(W_o[h_{t-1}, x_t] + b_o) \quad (9)$$

$$h_t = o_t * tanh(C_t) \quad (10)$$

The sigmoid function is one of them; $h_{t-1}$ represents the hidden state of the $t-1$ instant; $bb$ represents the bias of each gate; $i_t, f_t, o_t$ and $C_t$ are the input gate, forget gate, output gate, and unit status, respectively. For the connection, $W_f, W_i$ and $W_o$ are represented as a weight matrix. The three gates of LSTM cells govern the flow of information and hence define the cell's state. The gradient vanishing problem may be efficiently handled with LSTM [55].

The suggested model, in this instance, comprises two LSTM layers. The first LSTM layer outputs a sequence as one input above the LSTM layer. As explained previously, the internal design of both LSTM layers is the same. It also tried the LSTM cell rather than another GRU cell since the network with the LSTM cell outperformed the network with the GRU cell. This study constructs an LSTM network with a hidden vector of 128 elements. After the first LSTM layer, a dropout layer is added. Dropout reduces overfitting and enhances the model's generalization [56]. The LSTM's last layer generates a vector hi, which is supplied as the input to a fully linked multilayer network. Each layer has an activation function. The rectified linear unit (ReLU) activation function is used for each layer, and the exponential activation function is used for the output layer. Because the data set is binary, a nonlinear activation function is used to solve the binary classification issue. For hidden layers of neurons, the ReLU function was employed, while for the output layer of neurons, the sigmoid function was used.

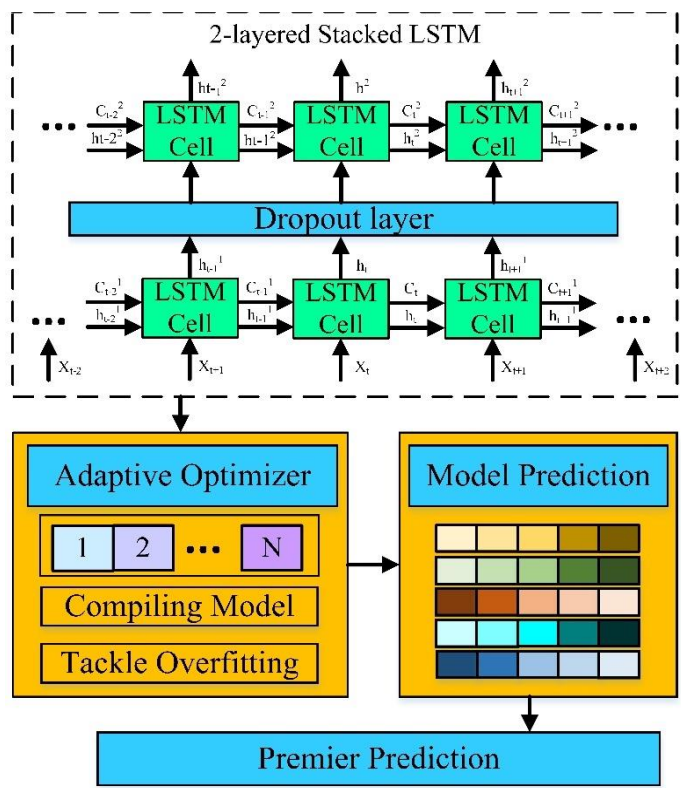

**Figure 4.** Phase II of the proposed stack generalization model.

After the training process, the parameters are changed or tweaked to assess the wrong predictions and ensure the predictions are correct as possible with optimization. Optimizer mold and design the model for the most accurate and possible prediction with the parameters (or weight). The value that the weights are updated in the training process is called the learning rate, a configurable hyperparameter to train deep neural networks with a small value within the 0.0 – 1.0 range. However, the learning rate varies due to overfitting [53]; thus, our model can predict accurately with the given dataset. Still, it is not appropriate for new or real-world data. We used the regularization technique to overcome the overfitting errors by fitting the functions appropriately on the training sets. It helps to attain optimal optimization solutions. These optimizers modify the neural network's attributes, i.e., weights and learning rates, to improve the accuracy.

Thus, we have utilized the following five adaptive optimizers to generalize the LSTM networks to overcome the overall loss and improve the accuracy. The selection of these optimizers is also given below:

- **AdaDelta:** This optimizer is based on the learning rate per dimension to address instead of the learning rate by parameter. It can solve the continual decay of learning rates by training and based on manually selected learning rates.
- **Adam:** It utilizes the prediction of the first and second moments to adapt the learning rate for the neural networks. It uses the momentum concept for adding a part of previous gradients to the current one. It is a faster optimizer and requires fewer parameters for tuning.

- **RMSprop:** Root means square propagation optimizer avoids the oscillations in the vertical direction and can increase the learning rate with feasible steps in the horizontal direction.
- **AdaGard:** It deals explicitly with individual features for different learning rates for different weights of sparse datasets to get a high learning rate. It can avoid the manual tuning of the learning rate for individual features.
- **SGD (Stochastic Gradient Descent):** Gradient descent optimizer has a drawback for large datasets. A variant of gradient descent, SGD, is generalized to make neural networks learn faster on the large-scale dataset.

These optimizers are implemented based on the packages and function calls in the Pytorch framework. For instance, we utilized $torch.optim.x$, where $x$ indicates the name of the optimizer, i.e., Adam or SGD, etc. The model will then be compiled using these adaptive optimizers. The model is trained to avoid overfitting by utilizing several epochs and early stopping strategies. By assessing the model using the test set, the output is now accessible. The stack generalization technique was used in the dataset after the strategy was implemented.

*3.4. Final prediction*

Two outputs are generated using the aforementioned multilayer stacked methods, and a model is chosen depending on the decision based on the value of the initial predictions. The mean prediction is considered and combined with the anticipated outcomes from the premier prediction. Finally, the outputs of the mean and premier prediction of the stacking models are combined as the final perdition using a meta-estimator classifier.

The meta-estimator involves constructing a robust classifier by applying the boosting method. Boosting combines multiple weak yet precise classifiers to create a powerful and resilient classifier for identifying phishing crimes. Additionally, boosting aids in integrating multiple features, resulting in improved classification performance. One notable boosting classifier is the XGBoost classifier, which transforms weak learners into potent contributors. It is well-suited for our proposed stack generalization model for identifying phishing sites, introducing a sense of symmetry to the classification process. Implemented on integrated feature sets of URLs and character-level features, it acts as a robust classifier within our proposed AntiPhishStack model for phishing identification, emphasizing the importance of symmetry in enhancing detection capabilities.

Suppose there are $n$ URLs in a set $\{(a_i, b_i) | i = 1,2, \ldots, n\}$, where $a_i \in E^f$; represents a set of selected features corresponding to $i$-th URLs while $b_i \in \{0,1\}$ is a class label, e.g., $b_i = 1$ if the URLs will be considered malicious or phishing websites. The final outcome of the XGBoost model will be computed using the following equation [57].

$$f_m(x) = f_l(b_i, f_m(a)) = \sum_{i=1}^{n} f_l\left(b_i, f_{m-1}(a_i) + G_m(a_i) + \Omega(G_m(a_i))\right) \quad (11)$$

where $f_m(x)$ is the model's prediction at step $m$, $f_l$ represents the training loss function, and $a$ represents the input features used in the XGBoost model. The regularization term $\Omega(G_m(a))$ is defined as $\gamma T + \frac{1}{2} \lambda \sum_{t=1}^{T} \omega_t^2$, where $T$ is the number of leaf nodes in the base learner $G_m(a)$, $\gamma$ is the complexity of each leaf, $\lambda$ represents the regularization parameter, controlling the strength of regularization in the XGBoost model, while $\omega_t$ is the output value at each final leaf node.

At step $m$, considering the base learners from previous steps $(m-1)$ as fixed, the loss function can be expanded using Taylor's series [57, 58]:

$$f_l(b, f_{m-1}(a) + G_m(a)) = \sum_{i=1}^{n} \left[g_i G_m(a_i) + \frac{1}{2} h_i G_m^2(a_i)\right] + \gamma T + \frac{1}{2} \lambda \sum_{t=1}^{T} \omega_t^2 \quad (12)$$

where $g_i$ and $h_i$ are the first and second derivatives of the loss function $f_1$ with respect to $f_{m-1}(a)$, computed as:

$$g_i = \frac{\partial f_l(b_i, f_{m-1}(a_i))}{\partial f_{m-1}(a)}$$

$$h_i = \frac{\partial^2 f_l(b_i, f_{m-1}(a_i))}{\partial f_{m-1}^2(a)}$$

This formulation defines the model's optimization process at each step, incorporating both the loss function and the regularization term to balance model complexity and fit. Then, the integrated features are categorized into phishing and benign, based on the weights by the meta-estimator for final prediction. Furthermore, XGBoost comes up with many advantages, some of which include (i) Within the training set, the power to fix missing values, (ii) working with extensive data that does not fit into memory, and (iii) to achieve the faster computing, XGBoost can utilize multiple cores on the CPU.

Deep learning involves many datasets and a significant time for model training. These models' efficiency depends on the system resource specifications and the complexity of datasets. In order to identify phishing assaults, the time complexity is a crucial factor [37]. The proposed method's computational cost is based on how the characteristics are generated and extracted. URL and character level features extracted by our proposed method require logarithmic time complexity $O(\log(n))$. The extraction of such features during the model training and time complexity depends on the number of samples $n$ and dimensions $d$. Accordingly, the time complexity of our proposed work is $O(n\log(n)d)$.

## 4. Experiments

This paper utilized Python 2.7 to develop the suggested model and TensorFlow GPU v1.8.0 as a machine learning framework. The operating system is Windows 10 Pro Education, and the architecture was built using Python. This project's Python packages and libraries to detect phishing URLs include *Keras* (built-in with *TensorFlow*), *SciPy*, *Pandas*, *NumPy*, *Matplotlib*, and *Seaborn*.

Support Vector Machine (SVM), Decision Tree (DT), Naive Bayes (GNB), Logistic Regression (LR), minimal sequential optimization. (SMO) and k-neighbor neighbors (KNN) algorithms were evaluated for stacking in this work. In the first stage, LSTM is employed as the basic classifier for stacked generalization, and further 10-fold cross-validation is utilized. In the Phase II, the XGBoost classifier is utilized as a meta-estimator for the final prediction.

For the model's effectiveness, the following statistical metrics are used to assess the proposed work for different purposes [59].

$$Accuracy = \frac{(TP + TN)}{(TP + TN + FP + FN)}$$

$$Precision = \frac{TP}{(TP + FP)}$$

$$Recall = \frac{(TP)}{(TP + FN)}$$

$$F\_measure = \frac{2 * precision * recall}{(precision + recall)}$$

Precision-Recall Curve: A graph is utilized for the trade-off between the true positive rate and the true negative or vice versa for the predictive model assessment [59].

- For Positive Precision $\quad P = \frac{TP}{TP+FP}$
- For Negative Precision $\quad N = \frac{TN}{TN+FN}$
- For Positive Recall $\quad PR = \frac{TP}{TP+FN}$

- For Negative Recall $\quad NR = \frac{TN}{TN+FP}$

where TP indicates the true positive, which means the number of URLs is correctly classified as phishing; in contrast, the parameter TN indicates the true negative, which means the number of URLs is correctly determined as benign. FP is a false positive, which means the number of benign URLs is wrongly classified as phishing, and FN is a false negative, which shows the number of phishing URLs classified as benign.

Mean Absolute Error (MAE): The average value for all absolute errors [60].

$$MAE = \frac{\sum_{i=1}^{n}|y_i - x_i|}{n} \qquad (13)$$

Mean Square Error (MSE): The average value for all squared errors [60].

$$MSE = \frac{1}{2}\sum_{i=1}^{n}(y_i - \hat{y}_i)^2 \qquad (14)$$

## 5. Results & evaluations

### 5.1. Feature evaluation with classifiers

This experiment evaluates both feature sets (URLF and CLF) from DS1 and DS2 by applying different machine learning classifiers. The fundamental goal of this experiment is to determine the best classifiers for both features in Phase 1. The future evaluations of these classifiers are given in Fig. 5 and Table 3. Imbalanced classes exist in real-time datasets, which creates the problem of classification. To tackle this problem, F-Measure is utilized due to the crucial values of false negatives and false positives. We refrained from using oversampling techniques such as the Synthetic Minority Over-sampling Technique (SMOTE) due to the potential risk of overfitting.

Figure 5 reveals a sense of symmetry, showcasing that SVM and NB exhibit maximum AUC, accuracy, precision, recall, and F-measure in DS1. These results show that SVM maintains an average accuracy of 91.1% and an 88.6% F-measure, while NB achieves an average accuracy of 87.295% and a 79.32% F-measure. However, some classifiers experience minimal accuracy. For instance, KNN records an average accuracy of 57.3% and a 51.25% F-measure, while LR demonstrates 67.46% accuracy and a 71.87% F-measure. The observed lower accuracy is primarily attributed to the independent protectors employed by these classifiers. Beyond the classifiers, the URLF feature in SVM symmetrically stands out with the highest accuracy, reaching 91.78%.

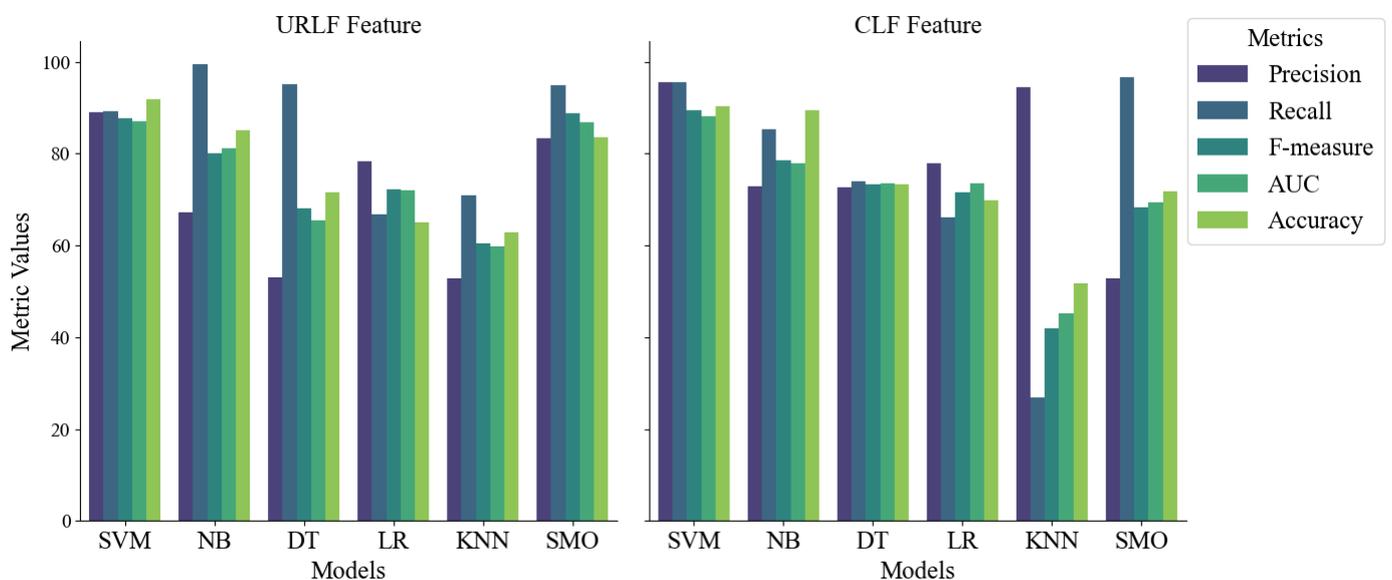

**Figure 5.** Classification results for proposed model on DS1.

Table 3 shows that the same classifiers have maximum accuracy to different extents in DS2. The results show SVM has an average of 92.09% accuracy and 93.64% F-measure, while NB has averaged 88.225% accuracy and 91.725% F-measure. Meanwhile, KNN and LR were found to have the lowest accuracy and F-measure. These lowest accuracies are due to the normal future distribution that is assumed by those classifiers.

**Table 3.** Classification results for proposed model on DS2.

| Models | Features | Precision | Recall | F-measure | AUC | Accuracy |
|---|---|---|---|---|---|---|
| SVM | URLF | 88.72 | 98.62 | 93.4 | 91.34 | 92.46 |
|  | CLF | 91.11 | 96.85 | 93.88 | 89.99 | 91.72 |
| NB | URLF | 86.26 | 98.18 | 91.84 | 91.07 | 90.08 |
|  | CLF | 86.03 | 97.96 | 91.61 | 78.68 | 86.37 |
| DT | URLF | 74.58 | 73.66 | 74.12 | 71.89 | 71.23 |
|  | CLF | 50.47 | 99.84 | 67.05 | 69.41 | 73.68 |
| LR | URLF | 80.61 | 99.92 | 89.23 | 85.37 | 76.24 |
|  | CLF | 69.62 | 81.25 | 74.98 | 74.02 | 71.82 |
| KNN | URLF | 78.24 | 66.93 | 72.14 | 70.56 | 70.34 |
|  | CLF | 78.38 | 66.86 | 72.16 | 71.21 | 68.91 |
| SMO | URLF | 86.03 | 97.21 | 91.28 | 91.4 | 89.64 |
|  | CLF | 78.42 | 98.68 | 87.39 | 87.24 | 83.21 |

As a comparative analysis of these features from both datasets (DS1 and DS2), Fig. 6 presents the bar graph, indicating the accuracy of the comparison of all machine learning used in this work. It can be easily distinguished that the classifier performances on DS2 are slightly better than on DS1 for the given features. However, some classifiers have minimum accuracy due to those features that are inefficient enough to discriminate between phishing and benign features. It might be possible that fishers are using modern technologies to receive online users and websites.

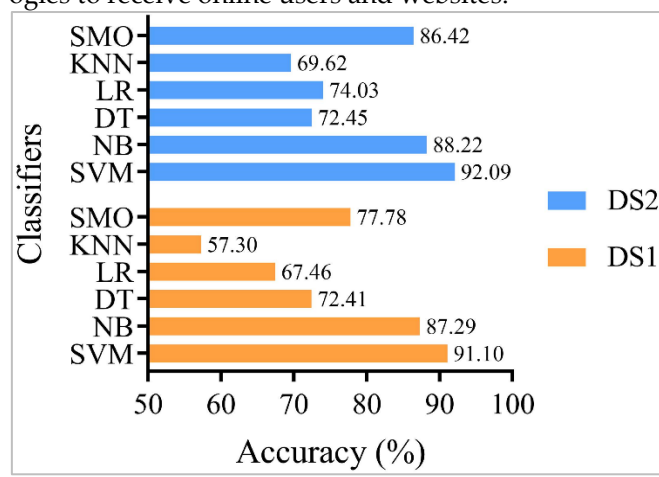

**Figure 6.** Comparison of machine learning classifier's performance for the proposed model on both datasets.

*5.2. Optimizer evaluation on LSTM*

Optimizers deal with model accuracy, which is a key component of machine learning and artificial intelligence, and it is responsible for molding the model to acquire possible accurate results. In this experiment, different levels of epochs used in machine learning and artificial intelligence are considered to indicate the number of passes of the entries to train the dataset. The different numbers of epochs are adjusted to implement the 2-layered LSTM with different optimizers. The sometimes higher number of epochs can lead to overfitting issues, and a lower number of epochs may result in underfitting the model. The learning rate controls the speed at which the mold learns. It is a configurable hyperparameter to train the neural networks with a small positive value from 0.0 to 1.0. From the previous experiment, it was detected that features from D2 have maximum accuracy and F-measure values. Therefore, the optimizers on both datasets are evaluated to confirm the result's originality with the LSTM-stack generalization model. Meanwhile, it used different deep learning-based adaptive optimizer programs from which the optimizer would be the best choice for the proposed anti-phishing model.

**Table 4.** Optimizer evaluations with LSTM-based stack generalization for DS1.

| Optimizer | Epochs | Features | Learning Rate | MSE | MAE | Accuracy |
|---|---|---|---|---|---|---|
| AdaDelta | 200 | URLF | 0.019 | 0.04 | 0.06 | 91.6 |
| | | CLF | 0.0023 | 0.08 | 0.05 | 91.1 |
| Adam | 100 | URLF | 0.007 | 0.09 | 0.07 | 89.3 |
| | | CLF | 0.0016 | 0.07 | 0.08 | 89.9 |
| RMSprop | 150 | URLF | 0.025 | 0.03 | 0.09 | 89.7 |
| | | CLF | 0.027 | 0.03 | 0.02 | 89.7 |
| AdaGard | 200 | URLF | 0.0048 | 0.06 | 0.03 | 91.8 |
| | | CLF | 0.098 | 0.02 | 0.02 | 92.5 |
| SGD | 250 | URLF | 0.003 | 0.07 | 0.09 | 90.4 |
| | | CLF | 0.003 | 0.05 | 0.06 | 90.2 |

**Table 5.** Optimizer evaluations with LSTM-based stack generalization for DS2.

| Optimizer | Epochs | Features | Learning Rate | MSE | MAE | Accuracy |
|---|---|---|---|---|---|---|
| AdaDelta | 200 | URLF | 0.0029 | 0.03 | 0.04 | 91 |
| | | CLF | 0.0017 | 0.06 | 0.05 | 90.6 |
| Adam | 100 | URLF | 0.096 | 0.07 | 0.08 | 91.3 |
| | | CLF | 0.0194 | 0.08 | 0.09 | 92.7 |
| RMSprop | 150 | URLF | 0.056 | 0.06 | 0.06 | 90.1 |
| | | CLF | 0.007 | 0.02 | 0.08 | 89.6 |
| AdaGard | 200 | URLF | 0.068 | 0.05 | 0.05 | 91.5 |
| | | CLF | 0.007 | 0.01 | 0.03 | 90.8 |
| SGD | 250 | URLF | 0.001 | 0.06 | 0.04 | 91.2 |
| | | CLF | 0.02 | 0.08 | 0.08 | 90.9 |

Table 4 shows that the AdaGard optimizer provider has the highest accuracy of 92.5%, minimum mean squared error (MSE), and mean absolute error (MAE) of 0.02 with CLF features in DS1. However, its learning rate is higher than other optimizers. Meanwhile, the performance of AdaDelta and the SGD optimizer is also significant, with the lowest learning rates of 0.0023 and 0.003, respectively. In contrast, the LSTM-based stack generalization model loses performance when predicting phishing features with other optimizers, such as Adam and RMSprop optimizers. Adam optimizer has an average of 89.6% accuracy and RMSprop 89.7 at both feature sets.

In a symmetrical comparison, Table 5 reveals that the Adam optimizer achieved a maximum accuracy of 92.7%, MSE of 0.08, and MAE of 0.09 with CLF features in DS2. It's noteworthy that the learning rate of the Adam optimizer is 0.0194. Simultaneously, AdaDelta, AdaGard, and SGD optimizers exhibit highly sufficient accuracy, MSE, and MAE while maintaining a minimal learning rate. However, RMSprop's performance is deficient, experiencing slightly lower accuracy with the proposed model, LSTM-based stack generalization. This decline stems mainly from the poor adaptive quality of those optimizers within the proposed model.

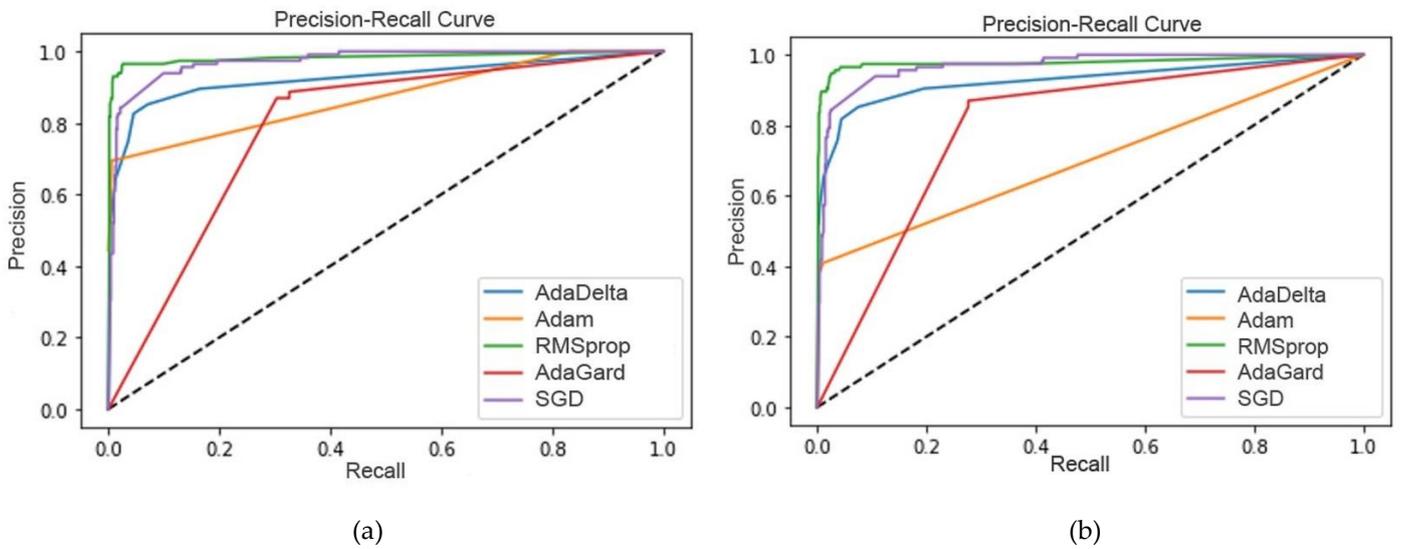

(a)                                                                  (b)

**Figure 7.** LSTM-based stack generalization model's Precision-Recall Curve on DS1 with (a) URLF and (b) CLF.

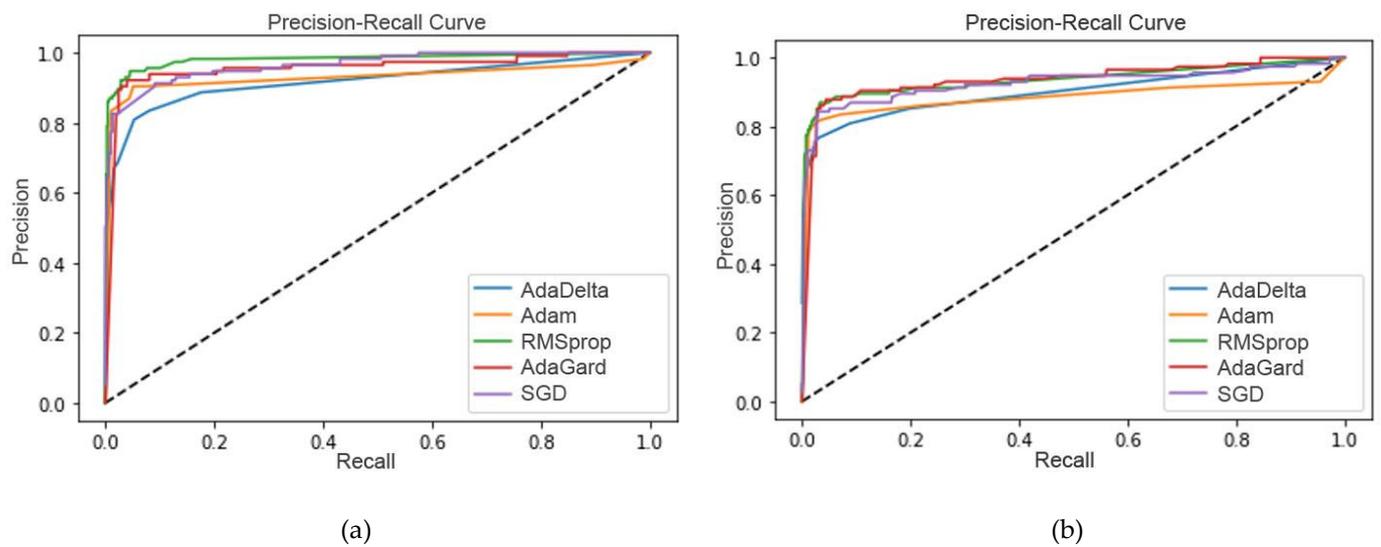

(a)                                                                  (b)

**Figure 8.** LSTM-based stack generalization model's precision-Recall Curve on DS2 with (a) URLF and (b) CLF.

Furthermore, precision-recall curves are illustrated in Figures 7(a) and (b) for each future. These curves indicate the trade-off between precision and recall. A big area under the curve shows high precision and recall; high precision indicates a low false-positive rate and high recall indicates a low false-negative rate. The analysis given in Tables 4 and 5 represents the learning rate that significantly contributes to the success of this proposed model with the adoptive optimizers. For example, the Adam optimizer has a maximum accuracy of 92.7, 0.08 MSE, and 0.09 MAE with CLF features for only a 0.0194 learning rate when the 2-layered LSTM is employed with 100 epochs. For comparing optimizers evaluation with LSTM-based stack generalization on both datasets, the average performance on DS1 is 90.62 accuracy, while 91.24 accuracies are reported on DS2.

In Fig. 7(a), the RMSprop optimizer has the highest precision-recall curve, while AdaGard has the lowest precision-recall curve with the URLF feature from DS1. Similarly, in Fig. 7(b), RMSprop also has the highest precision-recall curve, while Adam optimizer has the lowest precision-recall curve with the CLF feature from DS1. In Fig. 8, almost all optimizers have an equivalent precision-recall curve with URLF and CLF features from DS2. However, there is a slight difference between the performances of both future sets with the proposed model, LSTM-based stack generalization.

*5.3. AntiPhishStack model's evaluation*

The major goal of stacked generalization is to employ a next-generation-based model that combines the previous models to attain higher prediction-based accuracy. Generally, the stacking method merges the multiple models and learns them together for effective classification accuracy.

Initially, seven machine learning algorithms and one neural network have been harnessed on the given datasets to ensure the highest accuracy on both future sets. These algorithms play a pivotal role in crafting the proposed stack model. Furthermore, five widely recognized optimizers have been embraced to assess the LSTM-based stack generalization model, validating the symmetry between machine learning classifiers and a neural network. Consequently, the proposed model symmetrically classified and deduced phishing and benign features from both datasets. The classifiers and neural network-based optimization have already acquired symmetry in the stacking process. The AntiPhishStack model is constructed by amalgamating features from both Phase I and Phase II. Accordingly, the previous temporary prediction undergoes symmetrical filtering with a meta-estimator, XGBoost classifier, and LSTM model with its two hidden layers. Finally, the AntiPhishStack model deploys symmetrical detection criteria to distinguish the phishing and benign features.

For the performance evaluation of the AntiPhishStack model on given datasets, Table 6 presents a comprehensive measurement of the proposed model. The model training time is required to detect the features from the training dataset. The validation period presents the time required to classify each feature on the test datasets after the training is finished. Keras callback approach is used to note the model training and testing time. This approach can save the model performance time when accuracy is no longer improving and it interrupts the model on that particular task.

**Table 6.** Prediction of AntiPhishStack model with combined features on both datasets.

| Dataset | Training Time (s) | Test Time (s) | MAE | MSE | Precision | Recall | F-measure | AUC | Accuracy |
|---|---|---|---|---|---|---|---|---|---|
| DS1 | 8924.027 | 43.16 | 0.9 | 0.7 | 97.99 | 92.24 | 95.03 | 96.22 | 95.67 |
| DS2 | 9706.15 | 57.23 | 0.5 | 0.4 | 98.01 | 92.3 | 95.91 | 95.81 | **96.04** |

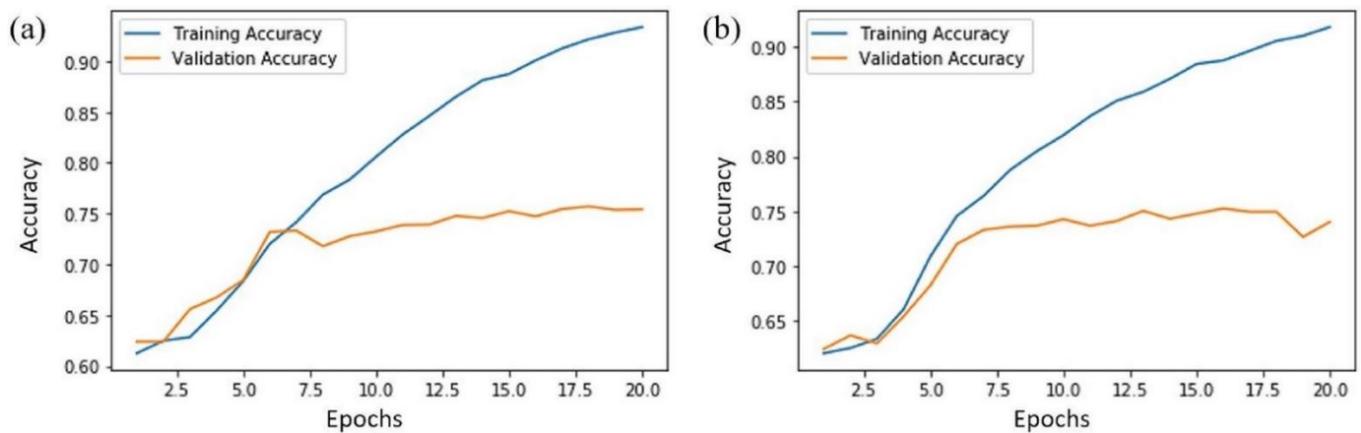

**Figure 9.** Training and validation accuracy of the AntiPhishStack model on DS1 with (a) URLF and (b) CLF.

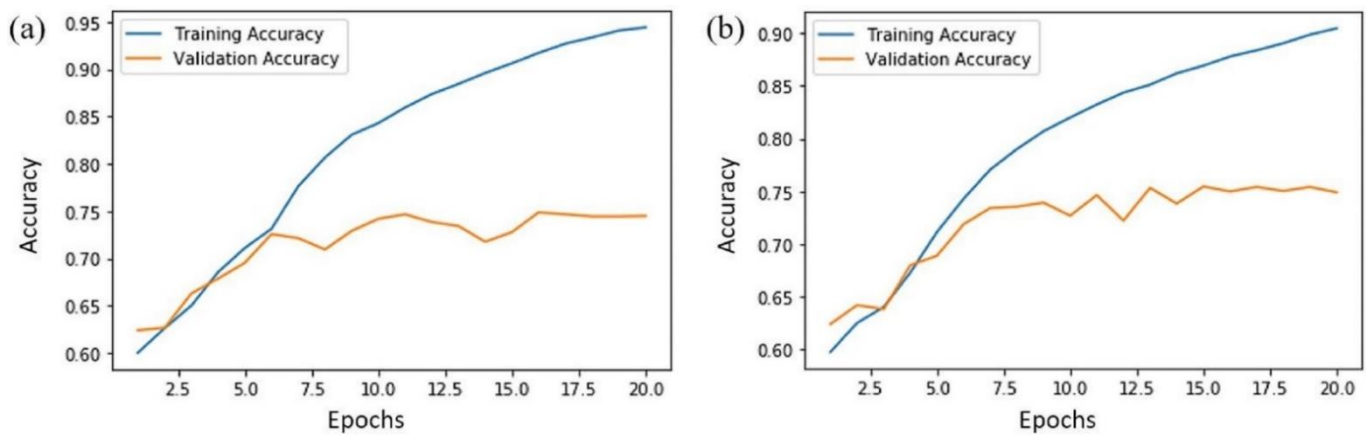

**Figure 10.** Training and validation accuracy of the AntiPhishStack model on DS2 with (a) URLF and (b) CLF.

It can be found that the AntiPhishStack model performed effectively on DS2 with 96.04% accuracy compared to DS1, which has slightly less accuracy. Similarly, DS2 has a better F-measure score and MAE and MSE rates. However, the training time for both datasets is unexpectedly higher. For example, this model provided phishing detection in 161 minutes for DS2 and 148 minutes for DS1. Although, in stacking models, the training time is always higher than in basic or hybrid models. This timing is higher in this case due to 2-layered LSTM and multiple stacks in both phases. Notably, in phases I and II, the average accuracy and F-measure values were less to a different extent than the final prediction with the AntiPhishStack generalization model. Furthermore, the training and validation accuracies have been provided with individual feature sets of each dataset. Fig. 9 presents the AntiPhishStack generalization model's training and validation accuracies for DS1, and Fig. 10 indicates DS2.

*5.4. Comparative analysis*

The significance of our proposed model has been described and illustrated in previous subsections. Furthermore, we have compared our model's performance with prior studies. Table 7 demonstrates the comparative analysis of existing studies by considering the applied approaches and algorithms WRT years.

It should be noted that there is some existing literature with higher performance than our proposed work; however, we have only compared those works that were nearly based on the stack generalization method, and these studies are tested in our same environment to ensure a fair comparison. The proposed model's performance outperformed the available studies to a different extent. For example, the current study [36] in 2021 proposed a generalized stack model for phishing detection by integrating URL features with CNN and XGB algorithms, and it reported 89.31% accuracy. In contrast, our proposed AntiPhishStack model received 96.04% accuracy by generalizing the URLs and character-level TF-IDF features.

**Table 7.** The proposed model's performance compared with the prior studies.

| References | Year | Applied approaches | Dataset | Algorithms/ networks | Observed accuracy (%) |
|---|---|---|---|---|---|
| [26] | 2018 | Web-based phishing using URLs features | UCI phishing dataset | Random forest | 92.9 |
| [21] | 2019 | BiLSTM for URL feature extraction and CNN for classification | Alexa and PhishTank | R-CNN | 95.6 |
| [46] | 2020 | Hybrid multimodel solution for networks | UNSW NB-15, UGR'16 | Ensemble four ML | 92.9 |
| [36] | 2021 | Integrated phishing URLs detection with CNN | PhishTank and Alex | CNN + XGB | 89.31 |
| Proposed model | 2024 | Stack generalized model for URLs & TF-IDF features. | Alexa, Yandex, PhishTank | LSTM and XGBoost | **96.04** |

*5.5. Limitation of the proposed model*

While our suggested method exhibits adequate accuracy, it does entail certain drawbacks. The first limitation is the asymmetry in our phishing detection method's dependence on English-language textual properties. In cases where the suspected webpage employs a language other than English, inaccurate classification findings may result. This study highlights the best-case performance of our method on multiple datasets, acknowledging the limitation of not providing an average performance metric across various conditions. Future research should incorporate a more comprehensive evaluation with rigorous average performance metrics.

Additionally, our methodology overlooks assessing the website's URL status, whether active or not, affecting overall outcomes. To surmount this limitation, expediting the training process and refining feature engineering becomes essential. Addressing zero-day assaults swiftly is imperative due to the narrow timeframe of phishing attacks. A pragmatic phishing detection system necessitates the inclusion of real-time detection as a crucial component. Capitalizing on big data technologies such as Apache Spark or Hadoop, which facilitate real-time processing, can reduce time complexity [61].

**6. Conclusion**

Phishing presents a symmetrical cybersecurity challenge. While machine learning techniques have played a pivotal role in phishing detection systems, recent advancements in deep learning yield significant outcomes for addressing contemporary phishing issues, especially with malicious URLs. This paper introduces novelty through a two-layered deep neural network (LSTM) integrated into a proposed stack generalization model that eliminates the need for prior feature knowledge in phishing detection. The model undergoes direct training on processed features, optimizing phishing URL detection through two phases. Results from DS1 indicate SVM's average accuracy at 91.1%, slightly improving to 92.09% in DS2. Furthermore, the AdaGard optimizer exhibits peak accuracy at 92.5%, with minimum MSE and MAE of 0.02 for CLF features in DS1. Comparisons with existing baseline models and traditional ML algorithms underscore the importance of the proposed model with its symmetrical stack generalization technique. Experiments demonstrate the AntiPhishStack model achieving 95.67% and 96.04% accuracy for the final prediction on benchmark DS1 and DS2, respectively. These results signify the model's intelligent detection of previously unidentified phishing URLs, positioning this paper as a symmetrical and advanced solution for establishing a phishing detection mechanism through profound deep learning-based stacking methods.

To detect malicious and fraudulent contract accounts, the AntiPhishStack model might be implemented with different deep neural networks, i.e., gated recurrent units (GRUs) networks with cryptocurrency-based features [11].

**Author Contributions:** Saba Aslam: Data curation, Writing - original draft, Writing - review & editing, Methodology, Software. Hafsa Aslam and Arslan Manzoor: Visualization, Writing - review & editing, methodology. Chen Hui and Abdur Rasool: Conceptualization, Supervision, Methodology, Formal analysis, validation. All authors have read and agreed to the published version of the manuscript.

**Funding:** This work is supported by Shenzhen Polytechnic Research Fund No. 6023310010K.

**Institutional Review Board Statement:** Not applicable.

**Acknowledgments:** The authors would like to thank all the anonymous reviewers for their insightful comments and constructive suggestions that have obviously upgraded the quality of this manuscript.

**Data Availability Statement:** The research data is used from https://yandex.com.tr/dev/xml/.

**Conflicts of Interest:** The authors declare that they have no known competing financial interests or personal circumstances that could have appeared to influence the work reported in this manuscript.